\documentclass[journal]{IEEEtran}
\usepackage{graphicx}
\usepackage{amsmath,bbm,amssymb,amsfonts,amstext,amsopn,sansmath}
\usepackage{cite}
\usepackage{balance}
\usepackage{url}
\usepackage{epsfig}
\usepackage{setspace}
\usepackage{stmaryrd}
\usepackage{psfrag}	
\usepackage{multirow}
\usepackage{makecell}
\usepackage{float}
\usepackage[process=auto]{pstool}
\usepackage{etoolbox}
\usepackage{algorithm}
\usepackage{algorithmic}
\allowdisplaybreaks
\usepackage{textcomp}

\usepackage{clipboard}
\newclipboard{Paper_V1}
 
\usepackage{xcolor}
\makeatletter
\let\myorg@bibitem\bibitem
\def\bibitem#1#2\par{%
	\@ifundefined{bibitem@#1}{%
		\myorg@bibitem{#1}#2\par
	}{%
		\begingroup
		\color{\csname bibitem@#1\endcsname}%
		\myorg@bibitem{#1}#2\par
		\endgroup
	}%
}
\newcommand{\highlightref}[1]{\expandafter\newcommand\expandafter*\csname bibitem@#1\endcsname{blue}}
\makeatother

\highlightref{zheng2019intelligent}

\newtoggle{OneColumn}
\toggletrue{OneColumn}

\newtoggle{arXiv}
\togglefalse{arXiv}

\usepackage{lipsum}
\def\@IEEEBIOphotowidth{1cm}    
\def\@IEEEBIOphotodepth{1cm}   
\def\@IEEEBIOhangwidth{1.2cm}    
\def\@IEEEBIOhangdepth{1.2cm}    

\EndPreamble
\begin{document}
\title{Intelligent Reflecting Surface-assisted Free-space Optical Communications}

\author{Vahid Jamali, Hedieh Ajam, Marzieh Najafi, Bernhard Schmauss, Robert Schober, and H. Vincent Poor \vspace{-0.5cm}%
	\thanks{V. Jamali and H. Vincent Poor  are with the Department of Electrical and Computer Engineering, Princeton University, Princeton, NJ 08544 USA (e-mail:  jamali@princeton.edu; poor@princeton.edu).}
\thanks{H. Ajam, M. Najafi, and R. Schober are with the Institute for Digital Communications at Friedrich-Alexander University Erlangen-N\"urnberg (FAU), Erlangen, Germany (e-mail:
	hedieh.ajam@fau.de;  marzieh.najafi@fau.de;		robert.schober@fau.de).}
\thanks{B. Schmauss is with the Institute of Microwaves and Photonics at FAU, Erlangen, Germany (e-mail:
	bernhard.schmauss@fau.de).}
}

\maketitle

\begin{abstract}
	Free-space optical (FSO) systems are able to offer the high data-rate, secure, and cost-efficient communication links required for applications such as wireless front- and backhauling for 5G and 6G communication networks. Despite the substantial advancement of FSO systems over the past decades, the requirement of a line-of-sight connection between  transmitter and receiver remains a key limiting factor for their deployment. In this paper, we discuss the potential role of intelligent reflecting surfaces (IRSs) as a solution to relax this requirement. We present an overview of existing optical IRS technologies; compare optical IRSs with radio-frequency IRSs and optical relays; and identify various open problems for future research on IRS-assisted FSO communications. 	
\end{abstract}

\begin{IEEEkeywords}
Intelligent reflecting surfaces, free space optical	systems, line-of-sight link, anomalous reflection, and relay nodes.
\end{IEEEkeywords}

\section{Introduction} 

Free space optical (FSO) systems are a promising candidate to support the high data-rate requirements of the next generation of wireless systems and beyond \cite{chowdhury20206g}. In particular, FSO systems can be deployed faster and more cost efficient compared to optical fiber links, while being able to offer data
rates on the order of several Gbps at a lower cost and less equipment weight  compared to their radio-frequency (RF) counterparts \cite{khalighi2014survey,kaymak2018survey}. In addition,  FSO systems are inherently secure and interference-free thanks to their narrow laser beams. These properties have made FSO systems an attractive option
for satellite, drone/balloon, and terrestrial communications particularly for wireless front- and backhauling  \cite{chowdhury20206g,khalighi2014survey,kaymak2018survey}.

\begin{figure}[t]
	\centering
	\includegraphics[width=0.98\columnwidth]{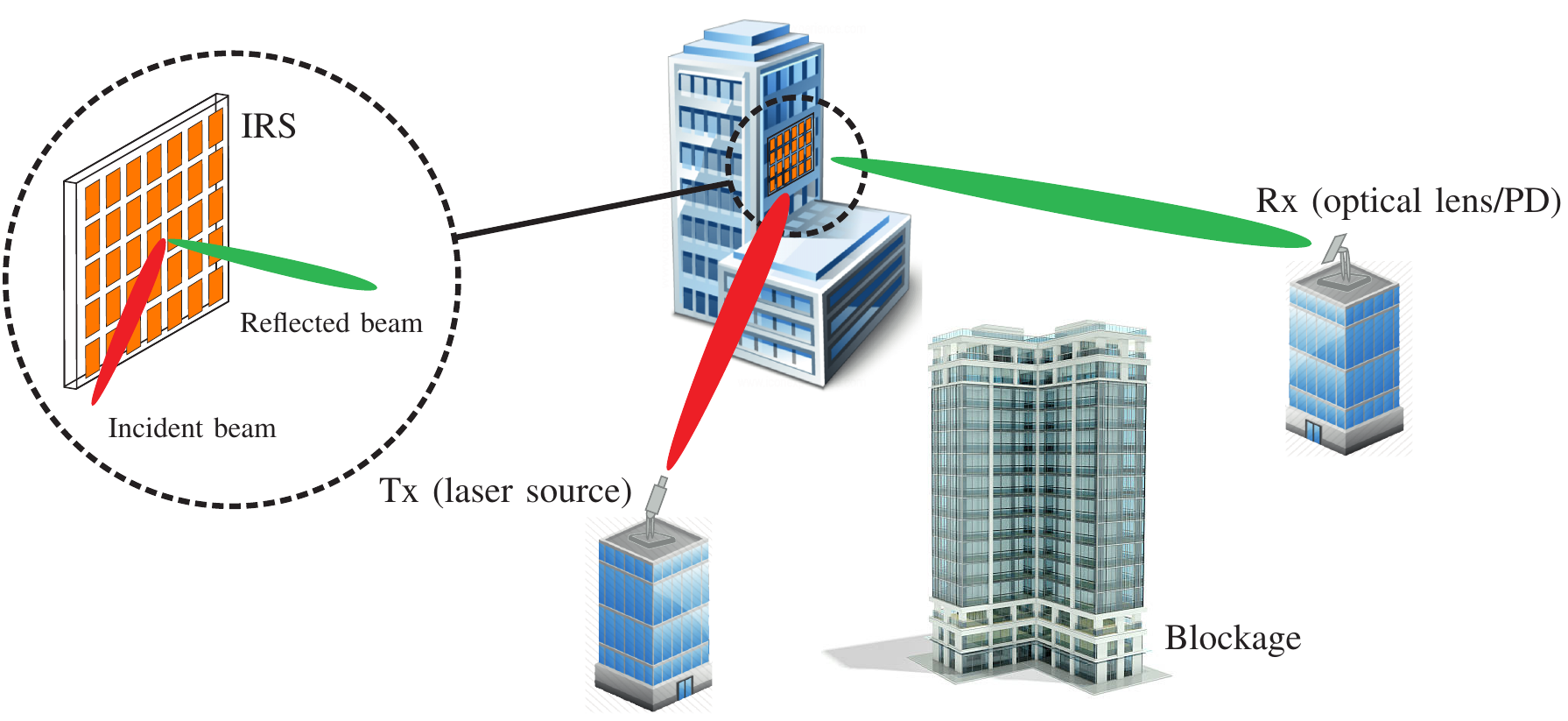}
	\caption{Schematic illustration of the deployment of an IRS to relax the LoS requirement between Tx (laser source) and Rx (optical lens and photo detector (PD)) in FSO systems by reflecting the incident beam on the IRS in the desired~direction.\vspace{-0.3cm}}\label{fig:system_model}
\end{figure}

Despite the aforementioned advantageous properties of FSO
systems, they face several challenges such as their susceptibility to atmospheric turbulence, pointing errors,
and high attenuation in adverse weather conditions. Over the past years, suitable countermeasures
have been developed to overcome these challenges including multiple-input multiple-output
(MIMO) FSO systems and hybrid RF/FSO systems \cite{khalighi2014survey}. However, the requirement of a line-of-sight (LoS) link between the transmitter (Tx) and the receiver (Rx) cannot be overcome with these techniques and constitutes a fundamental persisting limitation of FSO systems. Currently, the only viable solution for this problem is the deployment
of optical relay nodes. However, such relay nodes are expensive and inconvenient as they
require considerable additional hardware deployment. On the other hand, for RF communication systems, intelligent
reflecting surfaces (IRSs) have recently been shown to significantly improve the performance
of non-LoS wireless systems \cite{di2019smart,yu2021smart}. However, although there are few preliminary results \cite{najafi2019intelligent,najafi2020intelligentoptic,ajam2020channel,yang2020free,wang2020two,ndjiongue2021design}, the potential role of optical IRSs in the context of FSO communications (see Fig.~\ref{fig:system_model}) has not been well studied in the literature,~yet.   

The purpose of this article is to present a comprehensive overview of IRS-assisted FSO communications and identify the corresponding open research problems from a communication-theoretical point-of-view. To this end, we first introduce the main technologies available in the literature for the realization of optical IRSs and discuss their basic operating principles, their advantages, and limitations. Moreover, we review the important similarities and differences between optical and RF IRSs and compare the pros and cons of optical IRSs and optical relays. Furthermore, we present various potential directions for future research on IRS-assisted FSO systems including problems related to channel modeling, system design, performance analysis, and implementation. 

\begin{figure*}[t]
	\centering
	\includegraphics[width=0.95\textwidth]{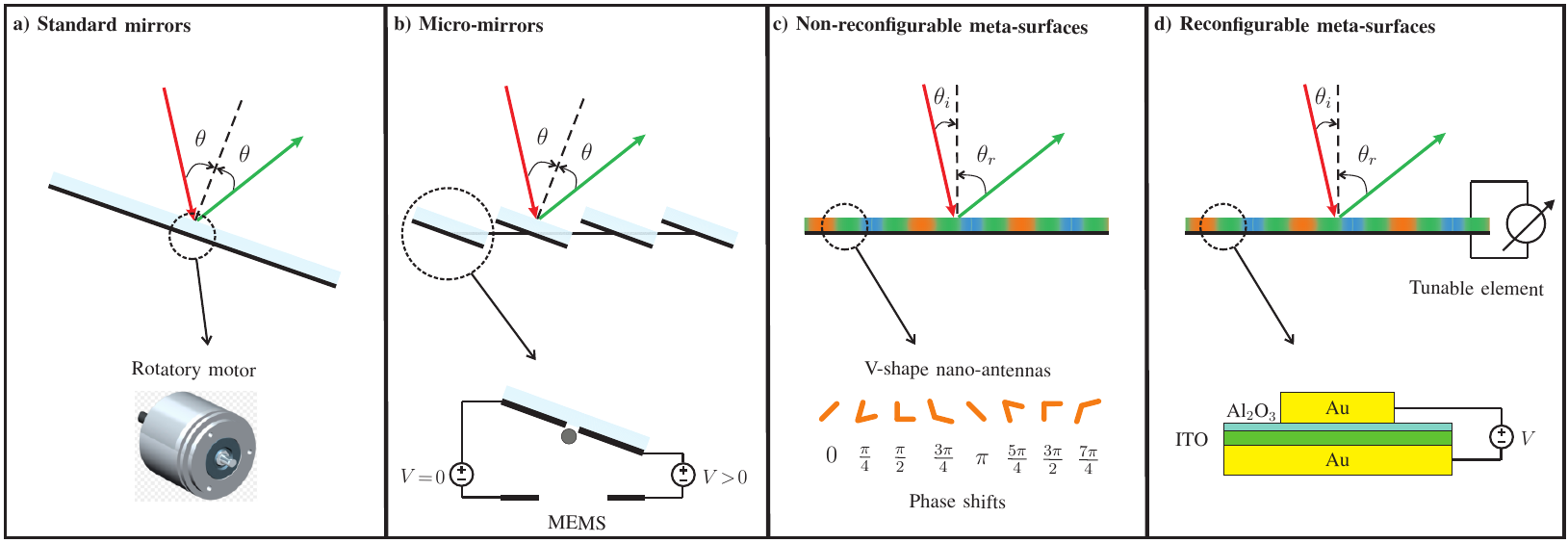}
	\caption{Schematic illustration of the considered optical IRS technologies and  exemplary options for their realization  \cite{najafi2019intelligent,chen2016review,wang2020mems,hail2019optical}. \vspace{-0.3cm}}\label{fig:system_IRSs}
\end{figure*} 

\section{Optical Reflecting Surface Technologies}

In this section, we review different technologies for the design of reflecting surfaces at optical frequencies.  We distinguish two general classes of optical IRSs, namely mirror-based and meta-surface-based systems. In the literature, some references categorize both spatial  light modulators \cite{goodman2005introduction} and optical phased arrays \cite{chen2016review} as optical meta-surfaces, whereas others distinguishes between them    \cite{hail2019optical}. Throughout this paper, we consistently use the term meta-surface to refer to surfaces which can be designed to locally (on the order of a wavelength) manipulate the incident electromagnetic wave including spatial light modulators.

\subsection{Mirror-based Designs}
Mirrors have been widely used in optical systems as reflecting devices \cite{goodman2005introduction}. Due to their limitation in supporting only specular reflection (i.e.,  the incident angle $\theta_i$ and the reflection angle $\theta_r$ are identical as $\theta_i=\theta_r=\theta$), mechanical change of their orientation  is needed in order to reflect the beam in a desired direction \cite{najafi2019intelligent}, which can be implemented as follows:

\subsubsection{Standard mirrors} If a single flat mirror is employed, the entire mirror  has to be re-oriented via, e.g., a mechanical rotary gimbal structure controlled by electrical motors, in order to reflect the beam in the desired direction, see Fig.~\ref{fig:system_IRSs}~a)  \cite{kaymak2018survey}. This design is cost efficient,  easy to analyze using geometric optics, and can serve as a baseline for the analysis of more sophisticated optical reflecting surfaces \cite{najafi2019intelligent}. Nevertheless, large mirrors require significant space for their re-orientation, i.e., a large form factor, which hinders their deployment,  e.g., on building walls. Moreover, the capabilities of mirrors are limited to specular reflection and no further modifications of the reflected beam are possible.

\subsubsection{Micro-mirrors} Micro-mirror surfaces, also referred to as deformable mirrors, consist of tiny mirror plates (whose dimensions are on the order of mm) mounted on micro-electro-mechanical systems (MEMSs) that enable their re-orientation/re-positioning, see Fig.~\ref{fig:system_IRSs}~b)  \cite{wang2020mems}. Due to the small size of each micro-mirror, the mechanical adjustment of the surface does not significantly increase its overall form factor, which implies that micro-mirror-based IRSs are approximately flat and hence suitable for deployment, e.g., on building walls \cite{wang2020two}. Moreover, the capability of individually controlling each micro-mirror can be exploited for new functionalities such as reflecting the optical beam incident on the surface in different directions realizing collimating and focusing functionalities. Nevertheless, micro-mirror surfaces are still constrained in their capabilities to manipulate the wavefront of the reflected~beam.  

\begin{table*}[t]
	\caption{Comparison of different optical IRS technologies discussed in this paper \cite{wang2020mems,chen2016review,hail2019optical}. Abbreviations: conductive oxide material (COM), phase change material (PCM), graphene (GRP), and liquid crystal (LC).}
	\label{table:IRScomparison}\footnotesize
	\centering
	\scalebox{0.85}{
	\begin{tabular}{||l|c|c|c|c|c|c||}\hline
		 \textbf{IRS technology} & \textbf{Basic operating physical principle}  & \makecell{\textbf{Control}\\\textbf{resolution}}&\makecell{\textbf{Operating}\\\textbf{wavelengths}} & \makecell{\textbf{Functional}\\ \textbf{capability}} & \textbf{Tuneability} & \makecell{\textbf{Technological}\\ \textbf{maturity}}\\
		\hline
	\textbf{Mirrors} & \makecell{specular reflection, \\ re-orientation of the entire mirror} &$\gtrsim 1$~cm& \makecell{near IR-visible}& low &low&very high\\
		\hline
	\textbf{Micro-mirrors} & \makecell{specular reflection, \\ re-orientation of micro-mirrors via MEMSs} & $\gtrsim 1$~mm&\makecell{near IR-visible}& moderate &moderate&high\\
		\hline
	\makecell[l]{\textbf{Non-reconfigurable} \\ \textbf{meta-surfaces}} & \makecell{change of geometrical properties of nano-antennas} & $\gtrsim 500$~nm&\makecell{GHz-visible}& high &none&moderate\\
		\hline
	\makecell[l]{\textbf{Reconfigurable} \\ \textbf{meta-surfaces}} &\makecell{variation of surface material properties \\ such as    charge density (e.g.,  COM and GRP), \\   structure of the material (e.g., PCM), \\ and molecular alignment (e.g., LC)}& \makecell{COM: $\gtrsim 1$~\textmu m\\ GRP:  $\gtrsim 10$~\textmu m\\ PCM: $\gtrsim 1$~\textmu m\\ LC:  $\gtrsim 10$~\textmu m}&\makecell{COM: near IR-visible\\GRP: THz-near IR\\PCM:THz-visible\\LC: GHz-visible}&high&high&low\\
		\hline
	\end{tabular}
}
\end{table*}


\subsection{Meta-surface-based Designs} 

Meta-surface-based IRSs consist of  a discrete planar array of subwavelength unit cells, which are able to  manipulate the properties of the reflected wave such as its phase, amplitude, and polarization. Although optical meta-surfaces are conceptually similar to RF meta-surfaces, the nano-scale technologies that are able to realize these  meta-surfaces at optical frequencies are quite different from the standard antenna technologies used for the RF bands. Therefore,  in the following, we review two main categories of optical  meta-surfaces and several implementation examples in each category in order to familiarize the reader with the corresponding state-of-the-art literature. 

\subsubsection{Non-reconfigurable meta-surfaces} Optical meta-surfaces were first designed based on the assembly of static nano-scale scatterers (i.e., nano-antennas) whose optical properties were invariant post fabrication \cite{chen2016review}. The basic underlying physical principle is that the properties of the reflected wave are a function of the physical geometry of the nano-antennas including their shape, size, and angular position. In particular,  the generalized  laws of reflection/refraction were first realized using
V-shaped optical nano-antennas in the mid-infrared spectral range by demonstrating that the full 360$^\circ$ phase-shift range can be realized by adjusting the angle between the legs of the V-shaped nano-antennas and their orientations, see Fig.~\ref{fig:system_IRSs}~c) 
\cite{chen2016review}. Examples of other phase-shifting mechanisms proposed in the literature include the variation of the size of isotropic metallic nano-patch antennas and angular rotation of  anisotropic scatterer elements to realize the so-called Pancharatnam-Berry phase change \cite{chen2016review}. Despite the non-reconfigurability of  these  meta-surfaces, they still may find applications in non-mobile FSO systems where the  transceiver positions are~fixed. 






%

\subsubsection{Reconfigurable meta-surfaces} Most reconfigurable meta-surfaces exploit materials whose optical properties can be modulated by an external electrical, mechanical, thermal, or optical stimulus \cite{hail2019optical}. In the following, we review two important  mechanisms enabling the tunability of optical meta-surfaces.

\textbf{Field-effect tuning:} Certain conductive oxide materials, such as  indium tin oxide (ITO) or indium zinc oxide (IZO), show a large change in their refractive index at near-infrared (IR) and even visible wavelengths if their charge carrier density is modulated via, e.g., an applied electrical field. For instance, the meta-surface in Fig. ~\ref{fig:system_IRSs}~d) consists of four layers, i.e., a metal-semiconductor-conductive oxide-metal  (Au-Al$_2$O$_3$-ITO-Au) configuration \cite{hail2019optical}. Upon applying a  positive (negative) voltage bias, the free charges of the ITO layer accumulate (deplete) at the semiconductor-ITO interface.
This variation of the charge accumulation layer causes a
significant change in the refractive index of the ITO layer which in
turn alters the phase of the reflected electromagnetic waves. Similarly, graphene has been shown to exhibit  significant charge carrier dependent modulation for THz to near~IR wavelengths if an electrical field is applied \cite{hail2019optical}.


%
%


\textbf{Structural change-induced tuning:} The so-called phase-transition materials are a class of materials that exhibit transitions between different states, which have distinct optical properties. For example,   vanadium dioxide (VO$_2$) switches between an insulator state to a metal  state when heated above a critical temperature, which can be achieved via the electrical Joule heating mechanism \cite{hail2019optical}. Similarly, upon thermal or optical stimuli, germanium-antimony-tellurium,   GeSbTe (GST), which is widely employed in rewritable optical disks, can switch between an amorphous (glass) state and a crystalline state with a pronounced change of the refractive index. Another tuning mechanism is based on liquid crystal (LC) materials (widely used in displays) whose molecular alignment structure can be controlled electrically, whereby different alignment structures produce different phase shifts \cite{ndjiongue2021design,hail2019optical}.

Table~I presents a comparative overview of the IRS technologies discussed in this section. We refer the readers to \cite{wang2020mems,chen2016review,hail2019optical} for more detailed comparisons.

\section{Comparison of Optical IRSs, Optical Relays, and RF IRSs} 

In this section, we discuss the key similarities and differences of optical IRSs compared to RF IRSs and optical relays, respectively.

\begin{figure*}[t]
	\centering
	\includegraphics[width=1.4\columnwidth]{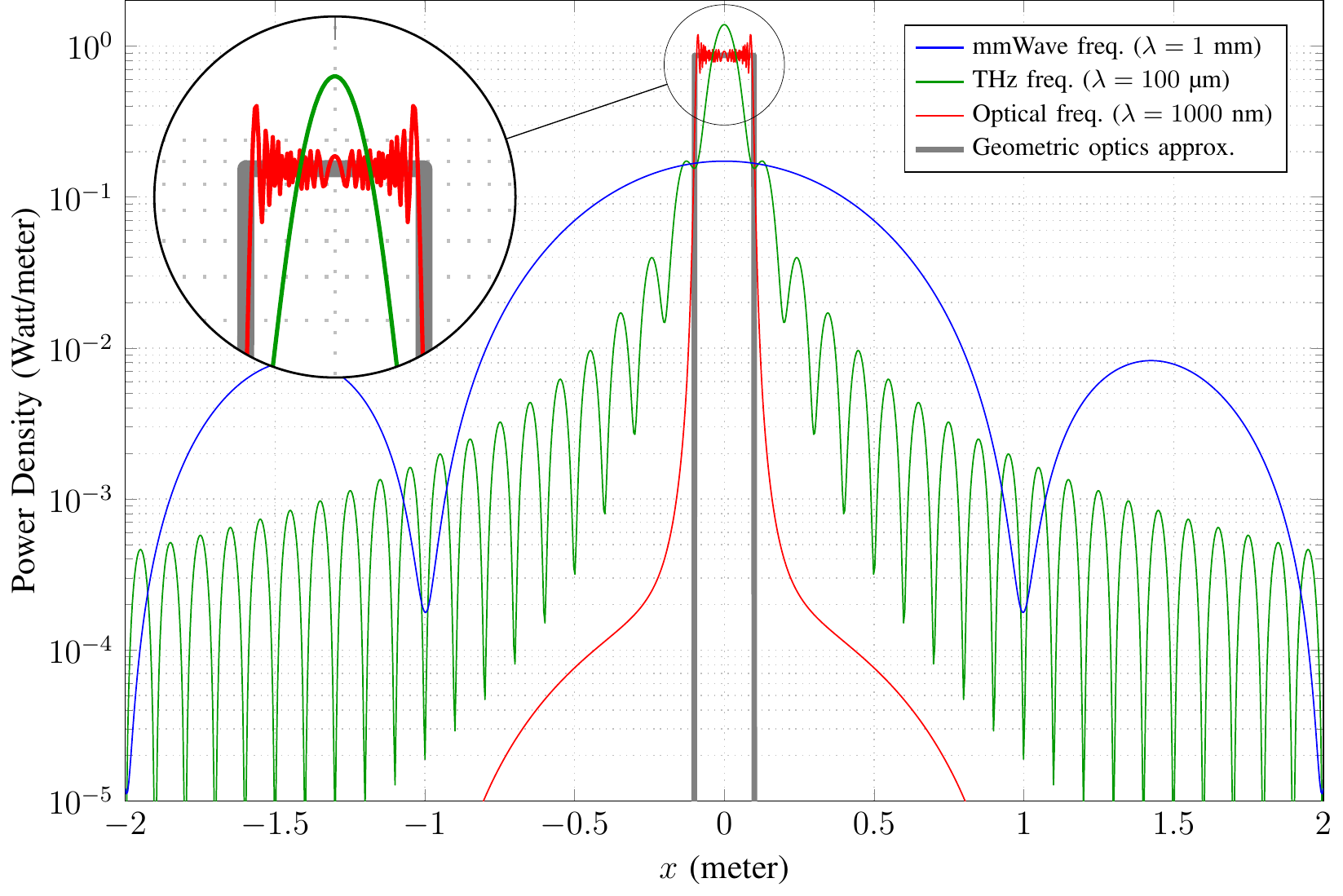}
	\caption{Power density of the reflected wave at $x$ and $y=200$~m. A 2D system is considered where an IRS of size $20$~cm is  located on the $x$-axis and centered at the origin. The IRS consists of  unit cells  with half-wavelength spacing and reflects an oblique  incident plane wave  ($\theta_i=30^\circ$ and a  transverse unit spatial power density) into the direction perpendicular to the IRS, $\theta_r=0^\circ$, using a linear phase-shift design \cite{ajam2020channel}. Apart from the curve obtained based on the geometric optics approximation, all curves are obtained using scattering theory, see \cite[Section~II-E]{najafi2020intelligentoptic}. \vspace{-0.3cm}}\label{fig:frequency_geometric}
\end{figure*} 

\subsection{Optical vs. RF IRSs}

While IRS-assisted RF systems  have been extensively studied  over the past few years \cite{di2019smart,yu2021smart},  IRS-assisted optical systems have remained relatively unexplored. In the following, we compare these two systems in order to help researchers   identify which results from the vast RF IRS literature  are applicable to optical IRSs (particularly optical meta-surfaces) and what are the distinct features of optical IRSs that require further investigation.

\subsubsection{IRS electrical size} The basic principle behind optical  and  RF meta-surfaces at wavelength level is essentially similar, i.e., changing the phase, amplitude, or polarization of the incident wave while reflecting it. However, in general, the  electrical dimension (i.e., the IRS dimension in meter divided by the wavelength)  of optical  IRSs is significantly larger than the electrical dimension of RF IRSs. An important consequence of a large electrical size is that tools from geometric optics, which are not applicable for the analysis of electrically small RF IRSs, become increasingly accurate for the analysis of optical IRSs. Fig.~\ref{fig:frequency_geometric} illustrates the example of anomalous reflection of a plane wave by a $20$~cm long IRS  at millimeter-wave (mmWave), THz, and optical frequencies. This figure shows that, for optical~IRSs, the result obtained based on the geometric-optics \cite{goodman2005introduction} (that models wave propagation by rays) agrees with the accurate result obtained based on the scattering theory \cite{najafi2020intelligentoptic} (that models each unit cell as a diffusive scatterer).

\subsubsection{Plane, spherical, and Gaussian waves} The optimization of  the phase shifts of the unit cells  of meta-surfaces  crucially depends on the phase  and power distributions of the incident wave on the IRS. For RF systems, the spherical wave propagation model is often used for near-field scenarios whereas the plane wave model is used  for  far-field scenarios \cite{di2019smart,yu2021smart}. Unlike these models, for optical systems, typically concentrated wave models, such as the so-called Gaussian beams, whose amplitude envelope in the  plane transverse to the wave propagation is a Gaussian function, are assumed \cite{khalighi2014survey,najafi2019intelligent}. This implies that increasing the IRS size beyond certain value does not significantly improve the performance because only the illuminated region of the IRS receives  significant optical power. Furthermore, beam tracking  becomes essential since the beam footprint may easily fall outside the IRS due to small pointing errors especially for large transmission distances.

\subsubsection{Power scaling law} 
\begin{figure*}[t]
	\centering
	\includegraphics[width=1.4\columnwidth]{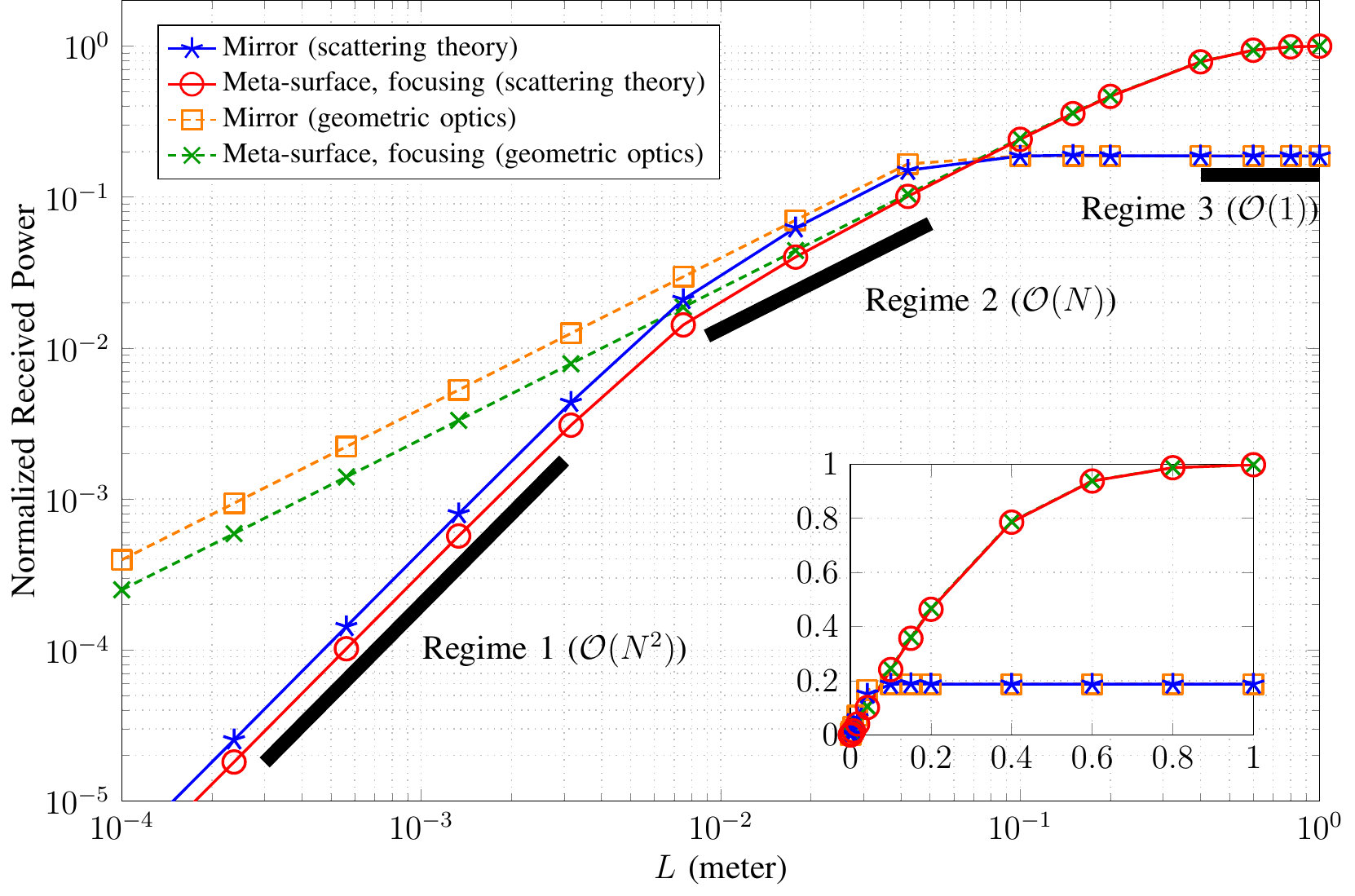}
	\caption{Fraction of transmit power received by an Rx lens vs. the IRS length $L$. We consider a 2D system, where the Tx, IRS, and Rx are located at $(-200~\text{m},300~\text{m})$, $(0,0)$, and $(0,500~\text{m})$, respectively. The Tx transmits a Gaussian beam at $1550$~nm wavelength with waist radius $w_0=1$~mm towards the IRS center. The IRS is located on the $x$-axis and has length $L$ and employs half-wavelength unit-cell spacing. The Rx lens has length $10$~cm,  is placed parallel to the $x$-axis, and~is assumed to collect all the incident optical power. Note that since the unit-cell spacing is fixed, the number of IRS unit cells $N$ grows linearly with the IRS size~$L$.  \vspace{-0.3cm}}\label{fig:scaling_law_area}
\end{figure*}

An interesting observation regarding RF IRSs is that for typical system parameters, the maximum power received at the Rx scales \textit{quadratically} with the number of IRS unit cells $N$, known as the quadratic scaling law $\mathcal{O}(N^2)$ \cite{di2019smart}, where $\mathcal{O}(\cdot)$ denotes the big-O notation. The quadratic scaling law originates from \textit{i)} the increased power received by the IRS, which scales as $\mathcal{O}(N)$, and \textit{ii)} the beamforming gain due to coherent signal superposition at the Rx, which also scales as $\mathcal{O}(N)$. We refer to the regime of the IRS size where  the quadratic scaling law is valid as Regime~1. For optical systems, both the IRS and the optical lens at the Rx are electrically very large such that the quadratic scaling law is not necessarily valid. Fig.~\ref{fig:scaling_law_area} shows the fraction of  the transmit power collected by an optical lens of length $10$~cm vs. the IRS length in a 2D system, where for the considered system parameters, the optical beamwidth on the IRS is approximately $30$~cm. For small IRSs in Regime~1, the Rx is in the far-field of the IRS, and hence, focusing becomes identical to anomalous reflection. Thereby, specular reflection by the mirror yields a slightly better performance than focusing (or anomalous reflection) by the meta-surface of the same size since the rotated mirror collects more power than the un-rotated meta-surface.
Moreover, as can be seen from Fig.~\ref{fig:scaling_law_area}, as the  IRS size increases, the power scaling  law transitions from quadratic $\mathcal{O}(N^2)$ to linear $\mathcal{O}(N)$, referred to as Regime~2.  This is because the Rx lens eventually collects almost all the power of the reflected narrow beam, and hence, the beamforming gain is not present any more. As the  IRS  size   increases further, the IRS collects  all the power sent by the Tx laser and hence the power received at the Rx lens saturates, too, referred to as  Regime~3. From Fig.~\ref{fig:scaling_law_area}, we  observe that  the geometric optics approximation is invalid for Regime~1 but becomes more accurate for Regimes~2 and 3, which correspond to practical optical systems.

\subsubsection{Channel impairments} The main channel impairments in RF communication systems are pathloss, small-scale multipath fading, and large-scale shadow fading. In contrast, FSO systems rely on the LoS link and therefore, multipath fading and shadowing do not exist. Instead, the FSO channel is influenced by \textit{i)} atmospheric loss which represents the power loss over a propagation path due to absorption and scattering of  light by particles in the atmosphere, \textit{ii)} atmospheric turbulence which is induced by inhomogeneities in the temperature and the pressure of the atmosphere, and \textit{iii)} geometric and misalignment losses which are caused by the divergence of the optical beam along the propagation distance and the misalignment between the laser beam line and the center of the Rx lens due to building sway. Among the aforementioned three factors, the geometric and misalignment losses of the end-to-end channel are most affected by the IRS \cite{najafi2019intelligent,najafi2020intelligentoptic}.

\subsection{Optical IRSs vs. Optical Relays}

Compared to IRSs, which only reflect a locally-modified wave from their surfaces, optical relays receive the signal, process it, and re-transmit it \cite{khalighi2014survey}. In the following, we discuss different considerations that have to be taken into account for a fair comparison of optical IRSs and relays. 

\subsubsection{Complexity} Optical relays require the entire transceiver architecture, e.g., a receiver lens, a transmitting laser, and depending on the relay type (e.g., amplify-and-forward (AF) or decode-and-forward (DF)), some analog and digital processing capabilities \cite{khalighi2014survey}. As discussed  in Section~II, the complexity of IRS-assisted FSO systems depends on the adopted IRS technology, which we may sort as standard mirrors, micro-mirrors, non-reconfigurable meta-surfaces, and reconfigurable meta-surfaces in ascending order of complexity. While mirror-based designs are certainly cheaper and simpler than optical relays, the complexity and cost comparison of optical meta-surfaces and relays is not straightforward, particularly due to the fact that optical meta-surfaces are not a commercially mature technology, yet. Nevertheless, since the considered meta-surfaces are passive (except for the tuning circuitry), they are expected to be in principle less complex compared with  optical~relays.

\subsubsection{Received power} For a fair comparison regarding the transmit power,  the sum of the transmit powers of the relay and the Tx laser sources in an optical relay system  should be set equal to the transmit power of the Tx laser source in an IRS-assisted system. In optical relay systems, the optical power that the relay and Rx collect depend on the transmit power, the beamwidth, and pointing errors.  In addition to these factors, in IRS-assisted FSO systems, the optical power that the Rx receives depends also on the IRS  size, which can be much larger than a receiving lens, and the IRS functionality, e.g., focusing in principle yields higher received power than anomalous reflection. Therefore, the received power in IRS-assisted systems can be larger or smaller than that in relay systems depending on the aforementioned parameter~values.

\subsubsection{Diversity gain} The variance of atmospheric turbulence is distance-dependent in FSO systems \cite{khalighi2014survey}. This leads to an improvement in the diversity gain of optical relay systems compared to the corresponding direct link  with the same overall distance. On the other hand, for IRS-assisted links, the variance of  atmospheric turbulence induced fading scales with the end-to-end distance. Assuming log-normal turbulence and plane wave propagation, the relative diversity gain of a two-hop equi-distant link compared to a direct link  is $2^{11/6}\approx 3.56$ \cite{khalighi2014survey}. Therefore, at high signal-to-noise ratios (SNRs), optical relay systems are expected to outperform optical IRS-assisted~systems.

\begin{figure*}[t]
	\centering
	\includegraphics[width=1.4\columnwidth]{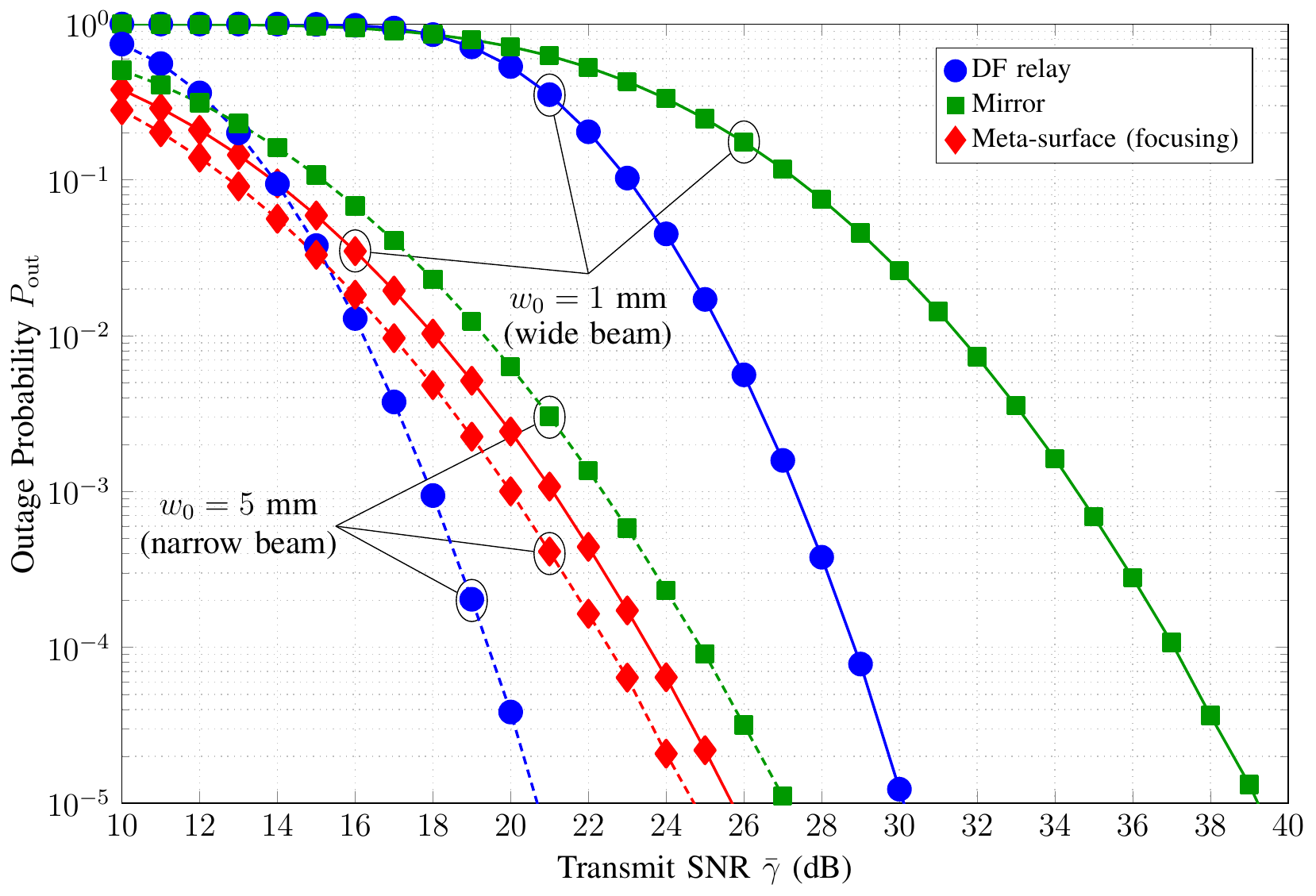}
	\caption{Outage probability $P_{\rm out}$ vs. transmit SNR $\bar{\gamma}$ for $\gamma_{\rm thr}=0$~dB and different beam waist radii $w_0$. The same setup as that in Fig.~\ref{fig:scaling_law_area} is considered. For the FSO channel, we consider weather-dependent attenuation with coefficient $\kappa=0.43 \times 10^{-3}$ m$^{-1}$, log-normal atmospheric turbulence with  index of refraction structure parameter $C_n^2=1.4\times 10^{-14}$, perfect beam tracking, IRS length $L=50$~cm, and photo-detector responsivity $\rho=0.5$, see \cite{najafi2020intelligentoptic} for details on the channel. The relay employs the same laser beam as the Tx and the same receiving lens as the Rx. Moreover, the transmit power is equally divided between the Tx and the relay. \vspace{-0.3cm}}\label{fig:relay}
\end{figure*}

Let $\gamma=\bar{\gamma}h^2$ denote the SNR, where $\bar{\gamma}$ is the transmit SNR and $h$ is the FSO channel coefficient, see \cite{najafi2020intelligentoptic} for a detailed characterization of $h$.
In Fig.~\ref{fig:relay}, we plot the outage probability $P_{\rm out}=\Pr\{\gamma<\gamma_{\rm thr}\}$ vs.   transmit SNR $\bar{\gamma}$, where $\Pr\{\cdot\}$ denotes probability and $\gamma_{\rm thr}$ represents the SNR threshold requirement. As can be seen from Fig.~\ref{fig:relay}, both mirror-based and  meta-surface-based IRS-assisted systems achieve SNR gains with respect to  optical relay-assisted systems for small $\bar{\gamma}$ while the SNR gain realized by   meta-surfaces is significantly higher than that realized by  mirrors. Nevertheless, due to their smaller diversity gain, IRSs are outperformed by optical relays at high $\bar{\gamma}$. Furthermore, Fig.~\ref{fig:relay} suggests that the SNR gain of  meta-surfaces over optical relays at low $\bar{\gamma}$ is more pronounced when the beam is wider. This is due to the fact that for wide beams, the relay and the Rx receiving lens cannot collect all of the optical power in relay systems, whereas the relatively large IRS can collect most of the optical power and efficiently focuses it towards the Rx lens.

\section{Future Research Directions}
In this section, we present various open research problems for IRS-assisted FSO systems.

\subsection{Channel Modeling}
The  system design and analysis of IRS-assisted FSO systems crucially depend on a sound understanding of how optical IRSs influence the FSO channel. Hence, we first discuss open problems related to channel modeling.

\subsubsection{Geometric loss} The amount of power collected by the Rx lens in an  IRS-assisted FSO system depends on the IRS orientation and/or specific phase-shift configuration  applied by the IRS. For anomalous reflection, the geometric loss was derived for mirror-assisted FSO systems  in  \cite{najafi2019intelligent} and  for  meta-surface-assisted FSO systems using a linear phase-shift  design in \cite{ajam2020channel} and a specific non-linear phase-shift design in \cite{najafi2020intelligentoptic}. The analysis of  the geometric loss  for other IRS phase-shift designs, e.g., for collimating or focusing, remains an open problem.

\subsubsection{Pointing error} Instead of targeting the Rx lens as in point-to-point FSO systems, in an optical IRS-assisted system, the laser source targets the IRS. Nonetheless, fluctuations of the center of the beam footprint on the IRS due to pointing errors translate into  a misalignment  of the beam footprint on the Rx lens plane, too. Moreover, the severity of the impact of pointing errors depends on the IRS phase-shift design and/or its orientation. For instance, while an IRS-assisted system designed for beam focusing may deliver a larger power to the Rx lens compared to a collimating IRS under perfect beam tracking, its performance can be more sensitive to pointing errors. Preliminary investigations of the impact of pointing errors on FSO channels assisted by mirror- and meta-surface-based IRSs can be found in  \cite{najafi2019intelligent,najafi2020intelligentoptic}.

\subsubsection{Channel delay dispersion} Unlike for specular reflection, for anomalous reflection, different points on the reflected wavefront may not travel the same distance. As a consequence, the transmitted signal gradually arrives at the Rx  up to a certain maximum delay which can be interpreted as an IRS-induced channel dispersion. The maximum channel delay for anomalous reflection of plane waves in a 2D system can be obtained as $D_{\max}=\frac{L}{c}|\sin(\theta_i)-\sin(\theta_r)|$, where $c$ is the speed of the light.  Such delay may cause inter-symbol interference (ISI) due to the large Gbps data rate of FSO systems, and necessitates the design of ISI-mitigating schemes at the Rx. For example, for $L=10$~cm, $\theta_i=0^{\circ}$,  $\theta_r=60^{\circ}$, and $10$~Gbps data rate, we have $D_{\max}=300$~ps and a symbol duration of $100$~ps, which implies each symbol is affected by ISI from the three previously transmitted symbols.


\subsubsection{Wavefront distortion} Practical IRSs are subject to phase-shift errors which can be caused by, e.g., a finite number of possible phase-shift  values, narrow-band phase-shifting  in wideband communication  systems, and diffraction effects due to the limited size of  IRS unit cells. The resulting phase-shift errors  cause a distortion of the wavefront of the reflected beam.
 In conventional FSO systems, optical techniques have been developed to correct wavefront distortions caused by atmospheric turbulence \cite{khalighi2014survey}. Further research is required to analyze the impact of IRS-induced wavefront distortions and to design efficient techniques to mitigate them.

\subsection{System Design and Performance Analysis}
In the following, we discuss several open research problems for the design and analysis of IRS-assisted FSO systems. 

\subsubsection{Initial link establishment} For meta-surface-based IRSs, the initialization of the IRS phase-shift configuration to ensure that a detectable power is received at the Rx is a challenging task. On the one hand, a random phase shift yields a negligible power at the Rx, and on the other hand, a desirable narrow beam may not fall into the Rx's field-of-view.  Even for point-to-point FSO systems, establishing the initial link is challenging \cite{kaymak2018survey}. One technique  to address this issue is to first generate a rather wide beam, whose footprint on the Rx plane may be on the order of several meters \cite{kaymak2018survey}. Once an initial connection is established, the IRS phase-shifts can be gradually refined to maximize the power received at the Rx. A similar design concept can be used for mirror-based IRSs based on optimizing the laser beam width and mirror/micro-mirror orientation.   Furthermore,  a backhaul control link is required to send feedback control signals from the Tx/Rx to the IRS, which can be realized by a low data-rate RF link.

\subsubsection{Channel estimation} The channel coherence time of FSO systems is on the order of $0.1$-$10$~ms, i.e., the channel stays constant over several thousand symbol intervals for typical data rates \cite{khalighi2014survey}. However, for  IRS-assisted systems,  only the end-to-end Tx-IRS-Rx channel can be estimated if the IRS is passive, which implies that   channel estimation and IRS  optimization are interleaved tasks. Due to the large number of optical elements in both meta-surface-based and  micro-mirror-based designs, a pragmatic channel estimation framework is to fix a given phase-shift/micro-mirror configuration at the IRS and estimate the end-to-end channel using conventional pilot-based techniques. Nevertheless, further research is required to study the efficiency of this approach or to develop more effective schemes.  

\subsubsection{IRS optimization} Depending on their type, IRSs can be optimized to realize various functionalities. For example, mechanically-tunable mirrors are able to steer the reflected beam in the desired direction; micro-mirrors can also realize beam collimating, focusing, and splitting; and meta-surfaces are in principle able to construct more complex shapes of the reflected wavefront. These capabilities can be exploited to maximize the optical power received at the desired Rx, to mitigate wavefront distortions, to simultaneously connect multiple Txs to multiple Rxs, etc. The optimization of the IRS for the aforementioned applications is an interesting future research topic.

\subsubsection{Optical information carriers} Commercial FSO systems often employ intensity/modulation direct-detection (IM/DD) systems where data is embedded into the intensity of the optical carrier \cite{khalighi2014survey}. For these systems, it suffices to design the IRS such that  maximum power is delivered to the Rx. However, information can be also embedded in the  phase and polarization of the optical wavefront \cite{khalighi2014survey}. In addition, data can be encoded into the orbital angular momentum (OAM) states of the light beam. For the latter systems, the IRS should be designed not only to  deliver a large power to the Rx but also to generate a reflected wavefront that is as un-distorted as possible. Concepts from adaptive optics can be used at the Tx and Rx in combination with IRS optimization in order to mitigate  phase wavefront distortion~\cite{kaymak2018survey,goodman2005introduction}. 

\subsection{Implementation}
While the focus of this paper is on the communication-theoretical aspects of  IRS-assisted FSO systems, we would like to highlight  some related implementation  aspects, as well.

\subsubsection{Relevant hardware constraints/impairments} While the optical IRS technologies reviewed in this paper have been experimentally demonstrated in the literature \cite{chen2016review,wang2020mems,hail2019optical}, their studied use cases are not the long-range IRS-assisted FSO systems considered here, and the designed IRSs, particularly in case of meta-surfaces, are often rather small, e.g., on the order of mm to cm \cite{chen2016review,hail2019optical}. Therefore, the development of experimental testbeds for IRS-assisted optical systems is of crucial importance as it helps to identify the relevant constraints and impairments in practical systems. 

\subsubsection{Verification of the theory} Theoretical models cannot fully capture the complex phenomena that are present in practical systems (even if they are known) and system designs are often based on first-order approximations of the actual systems. A critical question is whether these designs and performance analyses are valid for practical systems to an acceptable level of accuracy. Experimental testbeds are needed to verify the theoretical results and if needed refine them accordingly.  This is particularly true for systems involving new components such as optical IRSs.


\section{Conclusions}

In this paper, we have discussed the potential role of optical IRSs to relax the LoS requirement in FSO communication systems. We have reviewed different optical IRS technologies  and compared them in terms of  their operating principles, capabilities, and complexities. We have further compared IRS-assisted FSO systems with two related technologies, namely IRS-assisted RF systems and optical relay systems. Finally, we have presented an overview of various open research problems on channel modeling, system design, performance analysis, and implementation motivating future work on IRS-assisted FSO~systems.

\bibliographystyle{IEEEtran}
\bibliography{References}

\section*{Biography}

\noindent 
Vahid Jamali (Member, IEEE)  is a Postdoctoral Research Fellow at Princeton University, NJ, USA. 

\vspace{0.5cm}
\noindent 
Hedieh Ajam (Student Member, IEEE) is a Ph.D. student at Institute for Digital Communications (IDC) in  Friedrich-Alexander University (FAU) Erlangen-N\"urnberg, Erlangen, Germany. 

\vspace{0.5cm}
\noindent 
Marzieh Najafi (Student Member, IEEE) is a Ph.D. student at the Institute for Digital Communications (IDC) in  Friedrich-Alexander University (FAU) Erlangen-N\"urnberg, Erlangen, Germany. 

\vspace{0.5cm}
\noindent 
Bernhard Schmauss (Member, IEEE)  is a Professor for optical microwaves and photonics at the Institute of Microwaves and Photonics in the Friedrich-Alexander University (FAU) Erlangen-N\"urnberg, Erlangen, Germany. 

\vspace{0.5cm}
\noindent 
Robert Schober (Fellow, IEEE) is an Alexander von Humboldt Professor and the Institute for Digital Communications (IDC) in Friedrich-Alexander University (FAU) Erlangen-N\"urnberg, Erlangen, Germany. 

\vspace{0.5cm}
\noindent 
H. Vincent Poor	(Life Fellow, IEEE) is the Michael Henry Strater University Professor of Electrical and Computer Engineering at Princeton University, NJ, USA.

\end{document}